# A Solution to Bargaining Problem on Divisible Goods

Xuegong Deng

Northeastern university school of science, Shenyang, Liaoning, China 110004

Two-person bargaining problem is considered as to allocate a number of goods between two players. This paper suggests that any non-trivial division of goods cause a non-zero change on the solution of bargaining. So, a axiom of sharing division is presented, as an alternative axiom to Nash's axiom of independence of irrelevant alternatives and Kalai-Smorodinsky's axiom of monotonicity. This solution is targeted at the partialities of Nash and Kalai-Smorodinsky solution on some specific issues, but not to say it is better than others.

**Introduction**

This article discusses only the bargaining problems to allocate some goods between two players, the value of goods for the two players may be different. Goods are denoted by $X_k\{v_{k,1}, v_{k,2}\}$, $X_k$ is the name of the goods, and $v_{k,1}$, $v_{k,2}$ are the value of $X_k$ for player 1 and 2. In this article, the disagreement point is always $(0,0)$, so the goods sequence $X_k\{v_{k,1}, v_{k,2}\}$ can independently represent the bargaining. $\theta(X)=v_1/v_2$ is called the value ratio of X. If $X_k\{v_{k,1}, v_{k,2}\}$ can be divide into two parts, $X^1_k\{v^1_{k,1}, v^1_{k,2}\}$ and $X^2_k\{v^2_{k,1}, v^2_{k,2}\}$, and if the division satisfy

$$v^1_{k,1}+v^2_{k,2} > v_{k,1} \text{ or } v^1_{k,2}+v^2_{k,1} > v_{k,2} \tag{1}$$

which simultaneously means $\theta(X^i_k) \neq \theta(X_k)$, it is called a non-trivial division, otherwise it is called trivial division. After a non-trivial division, we get a new bargaining issue.

As the name has nothing to do with the allocation results, when we refer to N goods $X_k\{v_{k,1}, v_{k,2}\}$, the value ratio is always assumed in descending order. Then the upper right boundary of the utility sets is just the plotlines connecting the points $(\sum_0^n v_{k,1}, \sum_{n+1}^N v_{k,2})$. It has a little difference with the Pareto boundary, but we still call it Pareto boundary, since it will not cause any misunderstanding in this paper.

Since the correspondence between goods set and his Pareto set boundary, we can also call any segment in the plotline (Pareto boundary) as goods. If a point x is just a node of the plotlines, the value ratio of the left(right) segment is called left(right) value ratio of this point, and respectively denoted by $\theta^-(x)$ and $\theta^+(x)$. If the point is inside a segment, the left value ratio equals to the right value ratio, denoted by $\theta(x)$.

**Axiomatic definition**

The so-called bargaining solutions are all based on some axioms, so we can say that the solution is defined on the bargaining problems. The most famous axiom system is given by Nash, It consists of the following four axioms[1].

Axiom of Pareto optimality: Solution of bargaining is on the Pareto boundary.

Axiom of symmetry: Exchange the position of two participants does not affect the solution.

Axiom of invariance with respect to affine transformations of utility: Affine transformation on the feasible utility set (including the disagreement point) does not affect the solution.

The fourth axiom of independence of irrelevant alternatives is Nash's the most legendary axiom, and is also the most controversial one. If we denote a bargaining problem as a pair $(\alpha,S)$, in which $\alpha$ is the disagreement point, and S is the feasible utility set, and $f(\alpha,S)$ is the solution on the Pareto boundary, the axiom described as;

Axiom of independence of irrelevant alternatives: if $(\alpha,S)$ and $(\alpha,T)$ are two bargaining, $S \subset T$, $f(\alpha,T) \in S$, then $f(\alpha,S) = f(\alpha,T)$.

This solution can be simply explained as the point (x,y) on Pareto boundary, at where, $(x-\alpha_1)(y-\alpha_2)$ gets its maximum, $\alpha=(\alpha_1,\alpha_2)$. Nash solution has also a very intuitive geometric interpretation, see figure1.1. It happens to be the intersection of the Pareto boundary and the hyperbola, which centered at $\alpha$ and the intersection is one and only one point, see figure 1.1, point N. If the Pareto boundary is y=h(x), and Nash solution is (a,h(a)), then the Nash value ratio of this point is defined as $\theta^N=-h'(a)$.

Kalai-Smorodinsky questioned Nash's solution by very simple example (see figure1.2). Let $B_0$ denotes the bargaining $G_1\{75,25\}$, $G_2\{25,75\}$, $B_1$ denotes $G_3\{100,33.3\}$, $G_4\{0,66.7\}$ and $B_2$ denotes $G_5\{100,30\}$, $G_6\{0,70\}$, see figure1.2. The Nash solutions of $B_0$, $B_1$ and $B_2$ are respectively (75,75), (100,66.6) and (100,70). Kalai-Smorodinsky raised a sharper question.

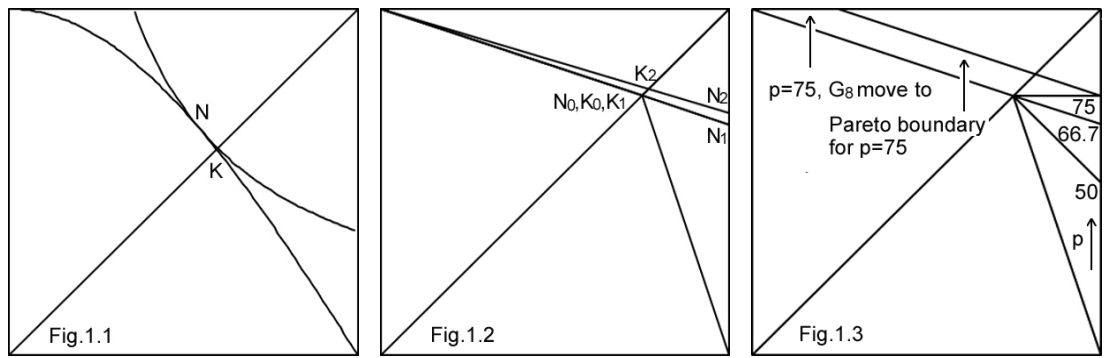

Figure1.1-3: 1. The sketch map of Nash and Kalai-Smorodinsky solutions. 2. The sketch map of $B_0$, $B_1$ and $B_2$. 3. The sketch map of $B_4$.

They asked that if the player1 get same income in $B_0$ and $B_2$, the player2 should get more in $B_2$ than in $B_0$, similarly, player1 should get more too if the player1 get same income in $B_0$ and $B_2$. So both player1 and player2 have a good reason to demand more in $B_2$ than in $B_0$. And in $B_1$, at least no one will reduce income compare to $B_0$. Both Nash solutions of $B_1$ and $B_2$ do not satisfy these demands, see figure 1.2, the Nash solution of $B_0$, $B_1$ and $B_2$ are marked as $N_0$, $N_1$ and $N_2$. So they suggested an alternative axiom for Nash's axiom of independence of irrelevant alternatives[2].

Kalai-Smorodinsky axiom of monotonicity: if ($\alpha$,S) and ($\alpha$,T) are two bargaining, their Pareto boundary denoted as y=$g_1$(x) and y=$g_2$(x), Max(x:(x,y)∈S)=Max(x,y∈T)), and $g_2$(x)≧$g_1$(x), then for the solutions (a,$g_1$(a)) and (b,$g_2$(b)) of this two bargaining, we have $g_2$(b) ≧$g_1$(a).

Like Nash solution, Kalai-Smorodinsky solution has also its geometry interpretation. If R is the smallest rectangle frame S, β is R's upper right corner, the solution of bargaining ($\alpha$,S) is just the intersection point of the Pareto boundary and the straight line connecting β and $\alpha$(see figure1.2, the three K-S solution points $K_0$, $K_1$ and $K_2$).

Let us return to the bargaining $B_0$, $\theta^-$ and $\theta^N$ are the left value ratio and Nash value ratio of (75,75), if $G_2$\{25,75\} can be divided into $G_7$\{0,p\} and $G_8$\{25,75-p\}, we get a new bargaining $B_4$. The change process of Nash, Kalai-Smorodinsky solutions with p are described in table1. When p>66.7, the segment $G_8$ will move to the most left of Pareto boundary, see figure 1.3.

Table 1. The change process of Nash, Kalai-Smorodinsky solutions with p, P=(100.p).

| Value of p | 0 to 50 | 50 to 60 | 60 to 66.6 | 66.6 to 75 |
| --- | --- | --- | --- | --- |
| N solution | unchanged | move to P | P | P |
| K-S solution | unchanged | unchanged | unchanged | Pareto improved |
| The value ratio of $G_8$ | $\theta(G_8)< \theta^N$ | $\theta^N<\theta(G_8)< 0.6$ | $0.6<\theta(G_8)< \theta^-$ | $\theta^-<\theta(G_8)< +\infty$ |

That is to say, for any division on the right of the solution, only when this division created a new segment G with $\theta(G)>\theta^-$, the Kalai-Smorodinsky solution get a Pareto improvement, otherwise, the solution is unchanged. Correspondingly, If $\theta(G)>\theta^N$, the income of player1 is improved in the Nash solution, but the new Nash solution may not be a Pareto improvement. This is the essence of the differences between Nash and Kalai-Smorodinsky solutions.

We cannot imagine a player would have no interest when facing a profitable division. So, this paper suggests that a non-trivial division must bring some player some benefit. This idea comes from the foregoing formula 1. We can do a simple reverse reasoning. Before the division, the bargaining have a solution, after a new division, there must be a new segment, the value for a player increased, so this player will hope to benefit from this division. So the author proposed a new axiom.

Axiom of sharing division: if $X_k$\{$v_{k,1}$,$v_{k,2}$\} is a bargaining, y=g(x) the Pareto boundary. (a,g(a)) is the solution of the bargaining under certain rule D, then the rule D can be described as: if a non-trivial division occurs on the right(left) of (a,g(a)), and if the new solution of the new bargaining on the new Pareto boundary y=$g^1$(x) is (b,$g^1$(b)), then b>a($g^1$(b)>g(a)).

The most intuitive meaning of this axiom is that, for any non-trivial split, at least one player will get benefits from it. So every player should not miss any possibility of profitable division, while trying to cover up the harmful possibilities.

This paper presents a solution to the Axiom of sharing division, it also satisfies the first three axioms. Y=g(x) is the Pareto boundary of a bargaining, if (a,g(a)) is the solution of this bargaining, and u=max(x), then it should satisfy

$$\int_0^a g(x)dx - af(a) = \int_a^u g(x)dx \tag{2}$$

Let us return to the bargaining $B_4$, we can see the difference among the three solutions. Figure 2.1-3 identified the three solutions in the case of p = 50, p=66.6 and p=75. In each figure, the two shaded area are equal, That is to say they satisfy the integral equation

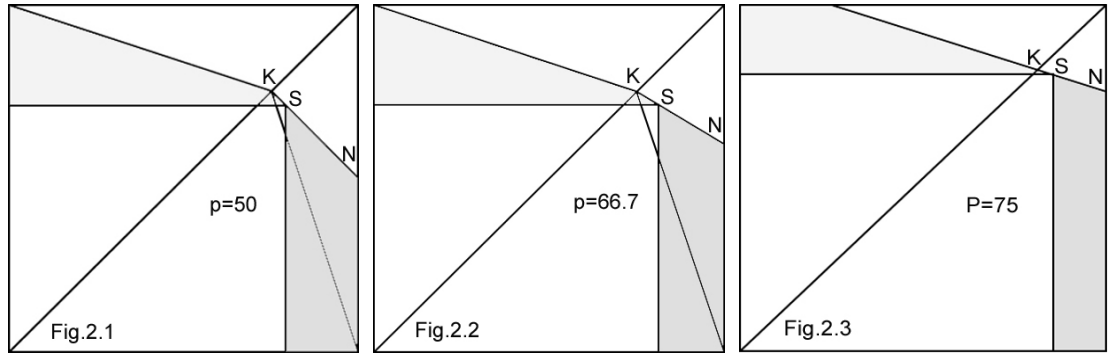

Figure2.1-3. comparison sketch maps for the three kinds of solutions.

**Discussion and conjecture**

The integral equation solution obviously satisfies the axiom of sharing division. This article conjectured that it is the unique solution to meet the axiom of sharing division. But authors have not got a strict proof. If it is not the unique solution, we need to find some more axioms to limit it.